\documentclass[twoside]{article}
\usepackage{fleqn,espcrc2}

\usepackage{graphicx}
\usepackage[figuresright]{rotating}

\newcommand{\AmS}{{\protect\the\textfont2
  A\kern-.1667em\lower.5ex\hbox{M}\kern-.125emS}}

\title{Spinodal Decomposition in Finite Temperature $SU(2)$ and $SU(3)$}

\author{Travis R. Miller \address[WU]{Department of Physics,\\
        Washington University,\\
        St. Louis, MO 63130, USA}
        and
        Michael C. Ogilvie \addressmark[WU]
\thanks{ We wish to thank the U.S. Department of Energy for financial support.}}

\begin{document}

\begin{abstract}
After a rapid increase in temperature across the deconfinement
temperature $T_d$ to temperatures $ T \gg T_d$, pure gauge theories
exhibit unstable long wavelength fluctuations in the approach
to equilibrium. This phenomenon is analogous to spinodal
decomposition observed in condensed matter physics, and also
seen in models of disordered chiral condensate formation. At
high temperature, the unstable modes occur only in the range
$0 < k < k_c$, where $k_c$ is on the order of the Debye screening
mass $m_D$. Equilibration always occurs via spinodal
decomposition for $SU(2)$. For $SU(3)$ at $T$ near $T_d$, nucleation
replaces spinodal decomposition as the dominant equilibration
mechanism.
Monte Carlo simulations of $SU(2)$ and $SU(3)$ lattice gauge theories
exhibit the predicted phenomena. For $SU(2)$, the observed value
of $k_c$ is in reasonable agreement with a value predicted from
previous lattice measurements of $m_D$.
\end{abstract}

\maketitle

\section{Introduction}

Much of the experimental relevance of finite temperature QCD
comes from non-equilibrium situations, as in the case of heavy ion
collisions. Current lattice techniques
are restricted to equilibrium simulations, and cannot address directly
non-equilibrium time evolution. However, there are aspects of non-equilibrium
behavior that are determined by the equilibrium effective potential.
After a rapid change in temperature, induced by some external agency, it
is the effective potential that determines the process of
re-equilibration, in particular whether relaxation, nucleation or
spinodal decomposition is the dominant equilibration mechanism.
We have shown that a large, rapid increase of temperature
across the deconfinement temperature results in spinodal decomposition
in pure $SU(2)$ and $SU(3)$ gauge theories.\cite{MillerOgilvie1}
In particular, we have found the existence of exponentially growing
long wavelength modes in the approach to equilibrium,
characteristic of spindodal decomposition.\cite{GuntonDroz}\cite{ChaikinLubensky}
This behavior depends only on the features of the
equilibrium effective action in an unstable region.

We consider first the generic case of a pure $SU(N)$
gauge theory in which the temperature is
raised rapidly from a temperature less than $T_{d}$, the
deconfinement temperature, to a temperature above $T_{d}$. 
We refer to such a rapid increase
in system temperature
as a quench, a borrowing of nomenclature from condensed matter physics.
The equilibrium order parameter for the deconfinement transition
is the Polyakov loop $L_F$.
At temperatures $%
T<T_{d}$, the $Z(N)$ global symmetry associated with confinement is
unbroken, and $\left\langle L_{F}\vspace{1pt}\right\rangle = 0$.
When the temperature is rapidly increased
to $ T > T_{d}$, $\left\langle L_{F}\vspace{1pt}\right\rangle \,=0$ is no
longer the stable state of the system, and must evolve to a new equilibrium
state with $\left\langle L_{F}\vspace{1pt}\right\rangle \,\neq 0$,
reflecting the transition to the gluon plasma phase.

\section{Effective Potential}

The instability of $\left\langle L_{F}\vspace{1pt}\right\rangle \,=0$ at
high temperatures is seen directly from the one-loop finite temperature effective
potential for the Polyakov loop, as derived by Gross, Pisarski and Yaffe and
Weiss.\cite{Gross}\cite{Weiss} For simplicity, consider the
case of $SU(2)$. 
It is convenient to parametrize the Polyakov loop as
\begin{equation}
L_{F}=\cos \left[ \pi \left( 1-\psi \right) /2\right].
\end{equation}
where $-1\leq \psi\leq 1$.
The effective potential at one loop for gauge bosons in a constant Polyakov
loop background is
\begin{equation}
V(\psi )=-\frac{\pi ^{2}T^{4}}{15}+\frac{\pi ^{2}T^{4}}{12}\left( 1-\psi
^{2}\right) ^{2}. 
\end{equation}
The one loop result dominates the effective potential for 
$T \gg \Lambda_{QCD}$ due to asymptotic freedom.
The first term is the standard black body result, obtained when $\psi = 1$.
The use of the variable $\psi $ makes the $Z(2)$ symmetry of the potential
under $\psi \rightarrow -\psi \,$manifest. Note that the equilibrium value
of $\psi $ is $\pm 1$, corresponding to $L_{F}=\pm 1$; $\psi =0$,
corresponding to $L_{F}=0$, is a maximum of $V(\psi )$.

\vspace{1pt}Our picture of the quenching process is that the system is
initially in a state where $\psi \,$is equal to zero at some temperature
below $T_{d}$. When the system is quickly raised to a new temperature $%
T>T_{d}$, the system is still in the state with $\psi =0$. However, the
system is unstable, and must eventually find its way to either $\psi =+1\,$%
or $\psi =-1$. Because we quench into a region of the phase diagram
where $V^{\prime \prime }(\psi )<0$, the system will
decay to the equilibrium state via spinodal decomposition.

Because pure $SU(2)$ gauge theory has a second-order deconfining transition,
spinodal decomposition will occur after quenching to any temperature $%
T>T_{d} $. The situation is more subtle for $SU(3)$,
where the deconfinement phase transition is first order.
This implies the existence of a metastable confined phase for
some range of temperatures above $T_d$, which in turn implies that nucleation
is the dominant equilibration mechanism just above $T_d$.
Perturbation theory gives no hint of this behavior, and
the one-loop effective potential is unstable at $L_F = 0$.
For temperatures
sufficiently large that the one-loop effective potential for $L$ is
reliable, spinodal decomposition will occur. 

\section{Langevin Model}

\vspace{1pt}In order to study the dynamics of this transition, we use the effective
action of Bhattacharya combined
with Langevin dynamics.\cite{Bhattacharya} 
For $SU(2)$, the effective action takes the form 
\begin{equation}
S_{eff}\left[ \psi \right] =\int d^{3}x\left[ \frac{\pi ^{2}T}{2g^{2}}\left(
\nabla \psi \right) ^{2}+ V(\psi)/T \right]. 
\end{equation}
The equilibrium distribution will be 
$ \exp \left[ -S_{eff}\left[ \psi \right] \right]$. 
We postulate Langevin dynamics of the form 
\begin{equation}
\frac{\partial \psi (x,\tau )}{\partial \tau }=-\Gamma \frac{\delta
S_{eff}\left[ \psi \right] }{\delta \psi (x,\tau )}+\eta (x,t) 
\end{equation}
where the white noise $\eta $ is normalized to 
\begin{equation}
\left\langle \eta (x,\tau )\eta (x^{\prime },\tau ^{\prime })\right\rangle
=2\Gamma \delta ^{3}(x-x^{\prime })\delta (\tau -\tau ^{\prime }). 
\end{equation}

The late-time relaxational behavior of $\psi $ is controlled by the Debye
screening mass $m_{D}$, given by $m_{D}^{2}=2g^{2}T^{2}/3$. Near
equilibrium, any effects of initial conditions decay exponentially as $\exp
\left[ -\frac{2\pi ^{2}\,\Gamma T}{g^{2}}(k^{2}+m_{D}^{2})\tau \right] $.
However, for early times, the initial conditions contribute to $\left\langle 
\widetilde{\psi }(k,\tau )\widetilde{\psi }(-k,\tau )\right\rangle $ a term
of the form 
\begin{equation}
\widetilde{\psi }(k,0)\widetilde{\psi }(-k,0)\exp \left[ -\frac{2\pi
^{2}\,\Gamma T}{g^{2}}(k^{2}-k_{c}^{2})\tau \right] 
\end{equation}
where $k_{c}^{2}$\vspace{1pt}$=g^{2}T^{2}/3=m_{D}^{2}/2$. Modes with $%
k<k_{c} $ are initially not damped but grow exponentially, with the $k=0$
mode growing the fastest. This is a characteristic feature of spinodal
decomposition with a non-conserved order parameter.
\cite{GuntonDroz}\cite{ChaikinLubensky}

\section{Monte Carlo Results}

\vspace{1pt}

\vspace{1pt}We have verified the existence of spinodal decomposition
with simulations of
rapidly quenched pure $SU(2)$ gauge theory. Lattices of size $32^{3}\times 4$
and $64^{3}\times 4$ were equilibrated at $\beta =2.0$, below the deconfinement
transition at $\beta _{d}=2.2986\pm 0.0006$.\cite{Fingberg}
The coupling constant was increased
instantaneously to $\beta =3.0$, and the approach to equilibrium monitored
via the Polyakov loop and other observables. The heat bath
algorithm was used both for equilibration at low temperature
before the quench and for subsequent dynamical evolution after the
quench. While
this time evolution is not the true time evolution of the
non-equilibrium quantum field theory, features such as spinodal decomposition
which depend only on the equilibrium effective action will occur
with any local updating algorithm which converges to the equilibrium
distribution.
The abrupt change in $\beta $ is a potential cause of concern with this
procedure, since the lattice spacing, and hence the
physical volume, changes in all directions when $\beta $
is changed.
However, the large spatial sizes used should
mitigate this effect. 

\begin{figure}[htb]
\vspace{-24pt}
\includegraphics[scale=0.32]{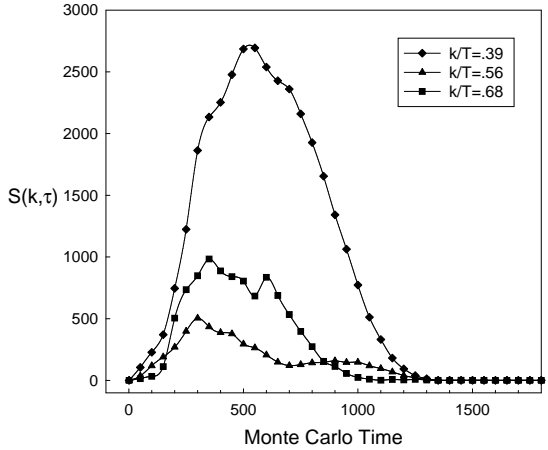}
\vspace{-40pt}
\caption{$S(k,\tau)$ versus Monte Carlo time.}
\label{fig:1}
\end{figure}
\vspace{-20pt}
Figure 1 shows the Fourier transform of the connected
Polyakov two point function $S(k,\tau )\,$for low values of the wave number
as a function of Monte Carlo time for the same simulation. Note the early
exponential rise in these modes, followed by a sharp disappearance as the
Polyakov loop reaches its equilibrium value, characteristic of spinodal
decomposition. Only the low momentum modes exhibit this growth; above $k_c$
no such growth occurs. Although the general behavior of $S(k,\tau )$ 
is the same for each run, many details are run dependent. In this
particular run, the $k/T = 0.68$ mode achieves a larger amplitude
than $k/T = 0.56$, which is atypical. In some runs, there is clear
evidence for mode-mode coupling, reflecting the nonlinearity of the system.
For each run,
we have estimated the rate of growth of each low-momentum mode by fitting
$log(S(k,\tau))$ to a straight line in $\tau$ for early times.
We can extract $k_c$ as the value where the growth rate is zero.
From equation $(6)$, the growth rate of each line is proportional to $k_c^2-k^2$
for the linearized theory, but this may not represent the true
time evolution of the simulation. In any case,
the growth rates measured in each run
are highly dependent on initial conditions.
In figure 2, we plot these growth rates versus $k^2/T^2$ for the same
run used in figure 1. 
The error bars are naive estimates of the error for each growth rate
for this particular run.
The x intercept
provides an estimate of $k_c^2$ for each run.
Using multiple $64^3 \times 4$ runs at $\beta = 3.0$, we have estimated $k_c/T$ to be
$1.14\pm 0.02$.
The principle errors in this estimate come
from the sensitivity of individual modes to initial conditions
and the discrete character of $k$ on the lattice.
We can compare this with lattice measurements of the Debye
screening length, assuming the one-loop relation
\begin{equation}
\frac{k_c}{T} = \frac{m_D(T)}{\sqrt{2}T}
\end{equation}
holds in general.
Using the results of Heller \textit{et al.}\cite{Heller} for $m_D(T)$, we obtain
$k_c/T =1.35(5)$. We consider this to be reasonable agreement,
given the many uncertainties involved.
\begin{figure}[htb]
%\vspace{9pt}
\vspace{-24pt}
\includegraphics[scale=0.32]{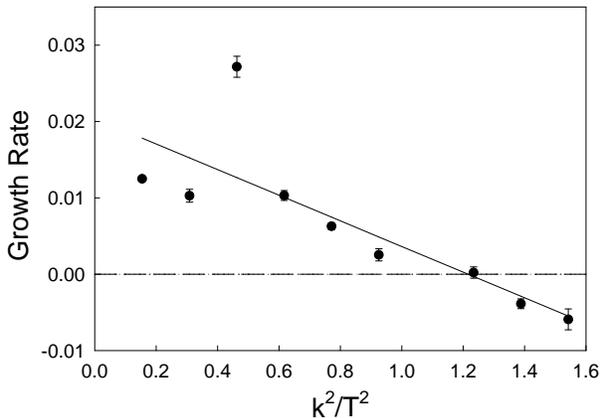}
\vspace{-40pt}
\caption{Growth rates versus $k^2/T^2$.}
\label{fig:2}
\end{figure}
\vspace{-20pt}

It is natural to ask what happens when rapid cooling takes place, as might occur
in the late stages of the expansion of a quark-gluon plasma, or in the early
universe. At low temperatures, we expect that $V(\psi )$ has a single
minimum and $V^{\prime \prime }(\psi )>0$ everywhere. Thus theory predicts the
absence of unstable modes, and that relaxational processes should dominate
the approach to equilibrium. Simulations of such a cooling
process, in which $32^{3}\times 4$ lattice configurations equilibrated
at $\beta =3.0$ are suddenly cooled to $\beta =2.0$,
shows no sign of spinodal decomposition.

We have begun studies of the more physically relevant case of $SU(3)$.
Since nucleation dominates at temperatures just above $T_d$, it is
necessary to go to higher temperatures to observe spinodal decomposition.
Working with $32^{3}\times 4$ lattice configurations equilibrated at 
$\beta = 5.5$, we performed quenches to various values of 
$\beta > \beta_d \approx 5.69$.
For values of $\beta$ close to $\beta_d$, 
we believe we see evidence for metastability.  
We have observed exponentially growing low wavelength modes for quenches to
$\beta \geq 5.80$.
\begin{figure}[htb]
\vspace{-24pt}
\includegraphics[scale=0.32]{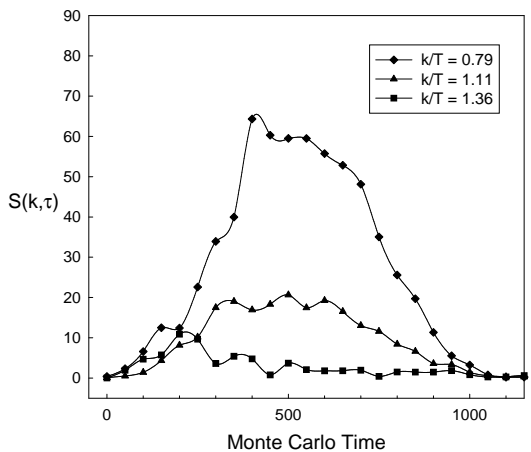}
\vspace{-40pt}
\caption{$S(k,\tau)$ versus Monte Carlo time.}
\label{fig:3}
\end{figure}
\vspace{-20pt}
In figure 3, we show the evolution of the three lowest non-zero modes as
a function of simulation time for a quench from $\beta = 5.5$ to $\beta = 5.92$.
Note the similarity to the behavior seen in
$SU(2)$. A systematic investigation is underway, emphasizing $k_c$ as well
as the limit of metastability.

\section{Conclusions}

Analytical and simulation results both indicate
the relevance of spinodal decomposition in the equilibration of a gluon
plasma after a rapid quench to high temperature.
In the case of $SU(N)$ gauge theories, we have shown that the physical
parameter $k_c$ which controls domain growth can be determined from
simulations. In the case of $SU(3)$, determination of $k_c$ as a function
of $\beta$ will allow us to define the limit of metastability as the point
$k_c(\beta_m)=0$. Note that small values of $k_c$ require correspondingly
large lattice sizes to see non-zero exponentially growing modes.

Just as the formation of a DCC\cite{DCC} may lead to enhanced production
of low-momentum pions, spinodal decomposition could lead to enhanced
production of low-momentum gluons in the early stages of plasma formation.
The characteristic scale for such a phenomenon would be $gT$.

Full QCD has quarks as well as gluons, and the phase structure
is different. It would be very interesting to study the behavior
of both the Polyakov loop and the chiral condensate
after abrupt changes in temperature.
The Polyakov loop can be studied fairly easily
if large equilibrated
unquenched lattice field configurations are available.
However, the chiral condensate requires a reliable estimator
for the local condensate as a function of space and simulation time.


\begin{thebibliography}{100}

\bibitem{MillerOgilvie1}
T. Miller and M. Ogilvie,
Phys. Lett. B488 (2000) 313.

\bibitem{GuntonDroz}
J. Gunton and M. Droz,
Introduction to the Theory of Metastable and Unstable States,
Springer-Verlag, Berlin (1983).

\bibitem{ChaikinLubensky}
P. Chaikin and T. Lubensky,
Principles of Condensed Matter Physics,
Cambridge University Press, Cambridge (1995).

\bibitem{Gross}
D. Gross, R. Pisarski and L. Yaffe,
Rev. Mod. Phys. 53 (1981) 43. 

\bibitem{Weiss}
N.~Weiss, Phys. Rev. D {\bf 24}, 475 (1981);
 Phys. Rev. D25 (1982) 2667.

\bibitem{Bhattacharya}
T. Bhattacharya, A. Gocksch, C. Korthals Altes and R. Pisarski,
Phys. Rev. Lett. 66 (1991) 998;
Nucl. Phys. B383 (1992) 497.

\bibitem{Fingberg}
J. Fingberg, U. Heller and F. Karsch,
Nucl. Phys. B392 (1993) 493.

\bibitem{Heller}
U. Heller, F. Karsch and J. Rank,
Phys. Rev. D57 (1998) 1438.


\bibitem{DCC}
K. Kowalski, J. Bjorken and C. Taylor,
SLAC-PUB-6109 (1993); 
K. Rajagopal and F. Wilczek, 
Nucl. Phys. B399 (1993) 395; Nucl. Phys. B404 (1993) 577.

\end{thebibliography}
\end{document}